\documentclass[review, times, authoryear]{elsarticle}

\usepackage{lineno, hyperref, gensymb, amsmath, cleveref }
\usepackage{amsfonts, mathtools, amssymb, longtable, enumitem, natbib, colortbl}
\modulolinenumbers[5]

\makeatletter
\def\ps@pprintTitle{%
	\let\@oddhead\@empty
	\let\@evenhead\@empty
	\def\@oddfoot{}%
	\let\@evenfoot\@oddfoot}
\makeatother

\newcommand{\be}{\begin{equation}}
\newcommand{\en}{\end{equation}}

\usepackage{etoolbox}
\patchcmd{\MaketitleBox}{\footnotesize\itshape\elsaddress\par\vskip36pt}{\footnotesize\itshape\elsaddress\par\parbox[b][36pt]{\linewidth}{\vfill\hfill\textnormal{\today}\hfill\null\vfill}}{}{}%
\patchcmd{\pprintMaketitle}{\footnotesize\itshape\elsaddress\par\vskip36pt}{\footnotesize\itshape\elsaddress\par\parbox[b][36pt]{\linewidth}{\vfill\hfill\textnormal{\today}\hfill\null\vfill}}{}{}%

\begin{document}
	
\begin{frontmatter}
\title{Daily Middle-Term Probabilistic Forecasting of \\
Power Consumption in North-East England}


\author[mymainaddress]{Roberto Baviera\corref{mycorrespondingauthor}}
\cortext[mycorrespondingauthor]{Corresponding author.} 
\ead{roberto.baviera@polimi.it}

\author[mymainaddress]{Giuseppe Messuti}
\ead{giuseppe.messuti@mail.polimi.it}

\address[mymainaddress]{Politecnico di Milano, P.zza L. da Vinci 32,  I-20133 Milan}

 \date{\today} 

\begin{abstract}
	Probabilistic forecasting of power consumption in a middle-term horizon (months to a year) is a main challenge in the energy sector.
	It plays a key role in planning future generation plants and transmission grid.
	
	We propose a new model that incorporates trend and seasonality features as in traditional time-series analysis and 
	weather conditions as explicative variables in a parsimonious machine learning approach,
	known as Gaussian Process.
	
	Applying to a daily power consumption dataset in North East England  provided by one of the 
	largest energy suppliers, we obtain promising results in Out-of-Sample density forecasts up to one year,
	even using a small dataset, with only a two-year In-Sample data.
	
	In order to verify the quality of the achieved power consumption probabilistic forecast 
	we consider measures that are common in the energy sector as pinball loss and Winkler score 
	and backtesting conditional and unconditional tests, 
	standard in the banking sector after the introduction of Basel II Accords.
\end{abstract}

\end{frontmatter}

	
	
\section{Introduction}

Power consumption forecast has received significant attention from both academics and practitioners in recent years.
In particular, 
middle-term forecast, i.e. in a time-horizon between a few months and a year\footnote{In the power consumption literature, short-term consumption forecasting is the 
prediction of the consumption (either of one operator or of the whole system) 
over an interval ranging from one day to a few weeks, 
while long-term consumption forecasting focuses on time-horizons longer than one year.}, 
plays a key role in the planning of power systems both for network reliability and for investment strategies in future generation plants and transmission \citep[see, e.g.,][]{HongFan2016}.

Through a probabilistic forecast, one obtains the full probability distribution of future consumption. It is the latest frontier of current research: 
it is more helpful for utilities and grid operators than point consumption forecast.
In fact, it does not provide only the expected value of the forecast but also information in terms of the dispersion 
of the forecast: a piece of information that is relevant for generating reliable scenarios.
This technique gained momentum in the energy sector after the Global Energy Forecasting competition 2014
(GEFCom) on a dataset of U.S.A. power consumption \citep[see, e.g.][and references therein]{HongFan2016, NOWOTARSKI20181548}.


In particular, 
it is important to be able to select the relevant drivers and their relationship with power consumption; 
this allows to understand and to hedge the risks.
We are interested in the impact of weather conditions; they play the most relevant role in middle-term forecasts compared to economic 
and demographic drivers that play a role in longer forecasts  \citep{HyndmanFan2010}.  
We focus on a region in the UK with relatively homogeneous weather conditions and on the power consumption of one main operator on the household energy market.
We desire to model the dependency of power consumption from weather conditions;
the most natural technique, now standard in power consumption forecasting, is known as \textit{ex-post} forecasting. 
It has been applied to middle-term probabilistic forecasting of power consumption on the French distribution network \citep{GNK2014} and 
on the National Electricity Market of Australia \citep{HyndmanFan2010}.

We use a Machine Learning (ML) technique. 
After the seminal paper of \citet{Park1991} these techniques have been shown to provide interesting results 
in short-term point consumption forecast
\citep[see, e.g.,][]{FanChen2006, Meng2009, Shi2018}. 
In this paper, we apply a ML technique also to a middle-term consumption forecast up to one-year; 
moreover, we consider a density  forecast and not only a point forecast.
Furthermore, the main successes of ML have been shown with big-data analytics \citep[see, e.g.,][]{Chen2012, Marino2016, Mocanu2016} and, 
in particular, they have been used in pricing forecast \citep[see, e.g.,][for a review]{NOWOTARSKI20181548}.
The real challenge is to use these techniques with small datasets.  
This is a common exigence in the industrial sector: it is quite hard to obtain within a middle-to-large operator homogeneous long time-series.
This fact is due both to the rapid changes that are observed in this energy market 
and to mergers and acquisitions, corporate transactions that have become frequent after the liberalization of the sector: 
the features of this energy market can change significantly over time the composition of clients' portfolio and then  the characteristics of the dataset.
In order to show the effectiveness and quality of the proposed technique,
we consider the extreme case where we analyse a two-year In-Sample dataset to forecast a one-year power consumption. 
Up to our knowledge,
the use of ML for middle-term density forecast using a small dataset is new in the literature.

In particular, 
the Gaussian Process technique is natural for density forecasting, because 
this model provides as output density forecasts of power consumption.
The model has a relatively low computational cost and it is very parsimonious - with only three parameters - compared to other ML techniques.
We are able to compute the densities of the consumption forecast for the hybrid model we introduce.

The performance of the model is obtained by comparing the forecasted results
over the last year of the dataset. 
In order to value the quality of the forecast we consider techniques that are standard either in the energy sector, as pinball and Winkler scores \citep[see, e.g.,][]{NOWOTARSKI20181548}, or in the banking sector after the introduction of Basel II Accords, 
as backtesting \citep{Kupiec1995, Christoffersen1998}

The main contributions of the paper are threefold.
First, we introduce a hybrid model that joints the advantages of
classical univariate time-series analysis and simple ML techniques. 
In particular, 
via a Gaussian Process we incorporate 
in power consumption density forecast the dependency from weather conditions and we deduce density characteristics for the hybrid model we consider.
Second, we show that  a ML technique relying on a small dataset can achieve promising results
forecast of power consumption using only weather data in middle-term forecasts.
Third, we value the density forecast via both sharpness and reliability measures, 
showing the quality of the achieved results.  
In particular, 
we show that a Gaussian Process can achieve very good results even considering a short daily time-series with only a two-year In-Sample set.

%

The rest of the paper is organized as follows. In Section \ref{sec:dataset} we summarise the key characteristics of the dataset we analyse. 
In Section \ref{sec:methodology} we present the methodology; in particular, we describe in detail i) the proposed model 
and how the weather conditions are introduced via a Gaussian Process, ii) the forecasting technique and iii) the evaluation methods. 
Section \ref{sec:results} shows the main numerical results and Section \ref{sec:conclusions} concludes.

\section{Dataset Description}
\label{sec:dataset}

North East England is one of the nine regions of England and the eighth most populous conurbation in the United Kingdom.
The dataset we analyse contains both the time-series of daily power consumption values and seven daily average weather conditions.
It is three years long, from April 2014 to March 2017.

Power consumption is the aggregated household consumption of one of the main UK power suppliers.
The weather dataset represents the daily average of hourly records of weather conditions in North East England. It includes seven different weather indicators:
	\begin{itemize}[noitemsep, nolistsep]
		\item Temperature, in $\degree C$;
		\item Wind Speed, in $m/s$;
		\item Precipitation Amount, in $mm$;
		\item Chill\footnotemark, in $\degree C \cdot m/s$;
		\item Solar Radiation, in $KJ/m^2$;
		\item Relative Humidity, in $\%$;
		\item Cloud Cover, on a scale 0 (clear) to 8 (completely cloudy).
	\end{itemize}
	\footnotetext{Chill is obtained combining temperature and wind speed; it represents how cooler one feels depending on the strength of wind \citep[see, e.g.][]{osczevski1995basis}.}

Table \ref{table:DescrStatsWeather} contains descriptive statistics about daily power consumption and weather data for the whole time window.
	\begin{table}  [h!]                                             
		\centering 
		{\footnotesize \begin{tabular}{|c|c|c|c|c|c|}                              
			\hline                                                        
			& Min & Max & Mean & Median & Standard Deviation \\                       
			\hline \hline     
			Consumption  [$MWh$] & 248.364 & 1208.705 & 632.921 & 638.327 & 249.996 \\ \hline \hline                                                         
			Temperature [$\degree C$] & -1.330 & 21.581 & 8.902 & 9.020 & 4.520 \\           
			\hline                                                      
			Wind Speed [$m/s$] & 0.667 & 10.219 & 3.400 & 3.000 & 1.624 \\            
			\hline                                                      
			Precipitations [$mm$] & 0.000 & 1.471 & 0.086 & 0.015 & 0.162 \\           
			\hline                                                      
			Chill [$\degree C \cdot m/s$] & 0.440 & 140.509 & 30.995 & 23.823 & 24.515 \\       
			\hline                                                      
			Solar Radiation [$KJ/m^2$] & 14.583 & 1212.917 & 397.424 & 327.083 & 296.357 \\  
			\hline                                                      
			Humidity [$\%$] & 60.167 & 100.000 & 85.415 & 85.344 & 7.525 \\    
			\hline                                                      
			Cloud Cover & 2.071 & 8.000 & 5.741 & 5.771 & 1.221 \\            
			\hline                                                      
		\end{tabular}        }                                       
		\caption{Descriptive statistics for daily power consumption and average daily weather data in NE England in the whole time window.}                                    
		\label{table:DescrStatsWeather}                                  
	\end{table} 
It is possible to notice that a wide variety of units of measurement in weather conditions:
it  implies ranges of values that can differ by orders of magnitude. 
Due to the difference in units of measurement, in order to obtain a homogeneous dataset it is often useful to standardise each input variable.
In the non-parametric  approach of  ML that we consider in the next section,  we use the Standardised Euclidean Distance.

\begin{figure*}[h!]
		\hspace{-1.2cm}
		\includegraphics[width=1.2\textwidth]{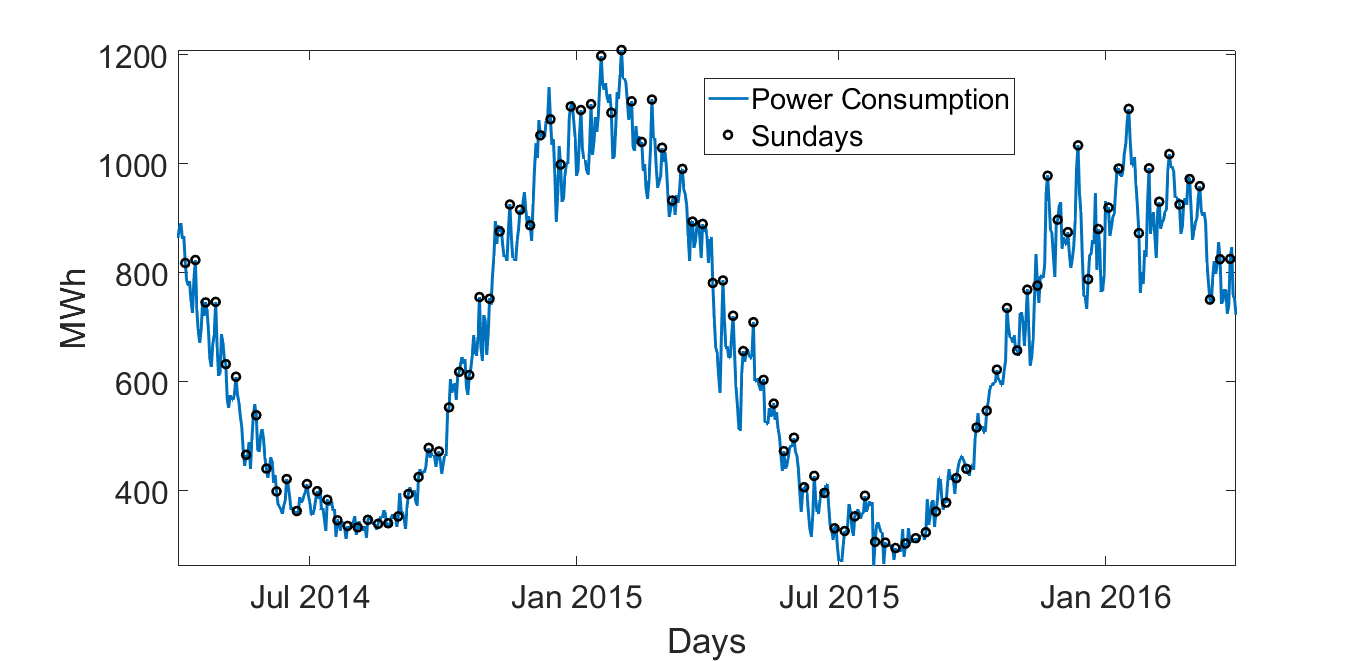}
		\caption{Power consumption (blue line) since April 2014 to March 2016. One can notice a yearly seasonal behaviour and locally higher values on Sundays 
(marked with a black circle).}
		\label{fig:powerconsumption}
\end{figure*}

The yearly and weekly seasonality of power consumption is rather evident in the dataset.
Figure \ref{fig:powerconsumption} represents the first two-year daily power consumption. 
It  suggests a yearly seasonal behaviour and slightly higher power consumption during weekends.
We observe that 
i) power consumption is  more than three times  larger in winter w.r.t. summer time, 
ii) consumption on weekdays is lower than 
 on Sundays (and to a lesser extent on Saturdays) and
iii) volatility is larger in winter than in summer time.
These are well known {\it stylised facts} common to most power consumption time-series: 
for these reasons, and in particular due to the observed volatility behaviour, it is now standard to model the logarithm of 
consumption \citep[see, e.g.,][for a review]{HongFan2016}.
In the next section we describe in detail the adopted methodology.

\section{Methodology}
\label{sec:methodology}

\subsection{The model}

The main goal of this study is forecasting 
future power consumption over the middle-term horizon, modeling both seasonal and weather-related features.
Our model choice is guided by the desire of obtaining a density forecast 
with a reasonably parsimonious description
without the need to resort to extremely fine-tuned models.

\bigskip

As already mentioned in the introduction, 
we model log-scaled daily power consumption data, as it is standard in the literature  for probabilistic electric consumption forecast.
The characteristics of power demand that we desire to model are:
\begin{itemize}[noitemsep,nolistsep]
	\item long-term trend;
	\item yearly and weekly seasonal behaviour;
	\item daily autocorrelation;
	\item the relation with weather conditions;
	\item weather-based error correlation and variance clustering.
\end{itemize}

The hybrid model we consider is split into two parts:
a classical approach that models the
first three characteristics (trend, seasonality and autocorrelation) and
a ML method that takes care of the weather influence.\\

First, the relation between consumption and calendar variables is established through a General Linear Model (GLM).
Then, we investigate the relation between GLM residuals and weather variables through a ML method known as a Gaussian Process (GP) \citep[see, e.g.,][]{Rasmussen06}.
We call this hybrid model GPX, because it is the natural extension  of AutoRegressive eXogenous models, 
known as ARX \citep[see, e.g,][p.534 et seq.]{box15}.

The model of the natural logarithm of power consumption can be written as:
\be
	Y_t = T_t + S_t + \gamma Y_{t-1} + R_t \quad ,
\label{eq:model}
\en
where
\begin{equation*}
		\begin{cases}
		T_t &= \beta_0 + \beta_1 t \\
		S_t &= \beta_2 \cos(\omega t) + \beta_3 \sin(\omega t) + \beta_4 D_{Sat}(t) + \beta_5 D_{Sun}(t) 
		\end{cases} \quad ,
\end{equation*}
where calendar time is measured via the cardinality $t$ of the observation,
starting from 1 on the first date in the dataset;
$T_t$ is the trend term, $S_t$ the seasonality both yearly and weekly, introduced via two dummy variables for Saturday and Sunday, 
$\omega = 2 \pi/365 $ and $ R_t $ are the residuals.
We also consider an Auto-Regressive (AR) term; 
in Section \ref{sec:results}
we show that the null hypothesis of a unit root can be refused  and
that an AR(1) describes properly the time-series. 

The described hybrid model approach where one differentiates first
trend, seasonality and AR components and then analyses separately the residuals  
 is standard  in the energy literature
\citep[see, e.g.,][and references therein]{Benth2008}. 
In this study residuals of power consumption
are modeled via a GP that  incorporates the information coming from weather conditions:
this is the main contribution of this study from a modeling perspective.

\bigskip


It is well known that, after having detrended and deseasonalized the time-series, 
the impact  on power consumption of weather conditions in general and of temperature  in particular, 
is very important and cannot be neglected \citep[see, e.g.,][]{HongFan2016}.

As already stated in the introduction, in this study 
we propose to incorporate weather conditions in the model via  a Gaussian Process.
GPs are well known in the ML literature and provide a simple tool for density forecast.
In this subsection we briefly recall the main characteristics of a GP, following the reference book of \cite{Rasmussen06} and using
a notation  similar  to the one they use in a {\it function-space view}.

A GP ``is a collection of random variables, any finite number of which have a joint Gaussian distribution" \citep[cf.][Def.2.1, p.13]{Rasmussen06}.
It  is completely specified by its mean function and covariance matrix.
In the case of zero mean, the random variables represent the value of the function $R(x)$ at ``location" $x$; 
in \citet{Rasmussen06} they are indicated as:
\be
R(x) \sim \mathcal{GP}\,(0, h(x, x')) \quad ,
\label{eq:GP_def}
\en
where $h(x, x')$ is an arbitrary kernel function between the locations $x$ and $x'$.

In practice, this notation indicates that for any collection of $n$ observations, 
the corresponding residuals are Gaussian random variables s.t.
\[
\mathbf{R} \sim \mathcal{N}(0, H(\mathbf{X}, \mathbf{X})) \quad , 
\]
where $\mathcal{N}( \cdot, \cdot)$ is a multinomial Gaussian distribution 
with zero mean and a positive definite covariance matrix  $H \in \mathbb{R}^{n \times n}$, $\mathbf{R} \in  \mathbb{R}^{n}$  and $\mathbf{X}  \in  \mathbb{R}^{n \times m}$, where $m$ 
is the number of regressors. 
In this study we consider $9$ regressors: 
the $7$ weather conditions in the dataset (cf. Section \ref{sec:dataset}) and $2$ other related to the calendar time $t$,
in order to introduce a yearly calendar effect also in the correlation matrix.
In particular, 
we consider $\cos (\omega t)$ and $\sin (\omega t)$
where $\omega = 2 \pi/365 $ is defined as in the seasonality $S_t$ in equation (\ref{eq:model}).
In the following, we continue to refer to these $9$ regressors as  weather conditions even if they contain these other two explicative variables.
The covariance $H_{ij}$ between the $i^{th}$ and $j^{th}$ residuals depends 
on the weather conditions $\mathbf{X}$ of the corresponding dates $t_i$ and $t_j$.

\bigskip
	
An example of kernel function is the Kronecker delta multiplied by a positive scalar ${\sigma}^2$, i.e.
$ h(x,x') = {\sigma}^2 \; \delta(x, x')$ ;
in this case, the residuals corresponds to Gaussian i.i.d. random variables with variance $ {\sigma}^2 $ as in the standard linear regression. 

In this study we consider a kernel function for the residuals equal to
\begin{equation}
	  h(x, x') = k(x, x')+{\sigma}^2 \; \delta(x, x') \quad ,
\end{equation}
where
\begin{equation}
	k(x, x') = \sigma_f^2 \, \exp\left({-\frac{\| x - x'\|}{\sigma_l}}\right)\label{eq:Exp} \quad ,
\end{equation}
with $\| x - x'\|$ the Standardised Euclidean Distance (SED) between $x$ and $x'$ and  $\sigma_f\ge0$, $\sigma_l>0$ two additional parameters w.r.t. 
the standard linear regression
\citep[see, e.g.,][]{lourenco2012short, morad2018electrical}.\footnote{In the ML literature a Gaussian Process is classified within the so-called non-parametric methods. 
This name does not indicate the absence of parameters, but the fact that the hypothesis space depends mainly from the dimension of the training set. 
The Gaussian Process we consider in this study involves three parameters: $\sigma_f$, $\sigma_l$ and ${\sigma}$.}
The choice of SED compared to L2 norm has the advantage of allowing the same contribution of each 
weather condition, independently from its unit of measure.
 
As standard in the statistical literature, the dataset is divided into 
In-Sample (IS) for model calibration (training set $X$) and Out-of-Sample (OS) for forecasting, with $n$ observations,
and evaluating the quality of the achieved forecast (test set  $X_{*}$), with a number of points equal to $n_*$, respectively two- and one-year long. 
One can compute pairwise, through the chosen kernel function, the covariance matrix of a finite number of GP observations 
\be
H(X,X) = K(X, X) + {\sigma}^2 I \quad ,
\label{eq:covariance}
\en
which depends from the three parameters $\sigma_f$, $\sigma_l$ and ${\sigma}$.

A GP presents the great advantage that it is immediate to infer the Out-of-Sample residuals and their distribution.
The distribution of the points in the In-Sample and the Out-of-Sample set is \citep[cf.][eq.(2.21), p.16]{Rasmussen06}
	\begin{align}
	& \begin{pmatrix*}[l]
	\mathbf{R} \\           
	\mathbf{R}_{*}
	\end{pmatrix*} \sim \mathcal{N}\, \begin{pmatrix}
	0,
	\begin{bmatrix*}
	K(X, X) + {\sigma}^2 I& K(X, X_{*})\\
	K(X_{*}, X) & K(X_{*}, X_{*})
	\end{bmatrix*}
	\end{pmatrix} \quad ,
	\end{align}
where 
$K(X, X_{*})$ denotes the $n_{} \times n_{*}$ matrix of the covariances evaluated at all pairs of training and test points, 
and similarly for the other entries $K(X_{*}, X_{*})$ and $K(X_{*}, X)$. 
Hereinafter, in order to simplify the notation, we indicate with $R_t$ the residual at time $t$ both IS and OS, where $R_t$ for $t=1,\ldots, n$ are the IS values
and  $R_t$ for $t=n+1,\ldots, n+n_{*}$ are the OS ones.

We can use the GP for the prediction of OS residuals. 
\citet{Rasmussen06} show that 
the OS residuals $\mathbf{R}_{*}$  given the IS residuals $\mathbf{R}$ and the weather conditions both IS, $X$, and OS, $X_{*}$, are (cf. eq.(2.22), p.16):
\[
	\mathbf{R}_{*} | X_{*}, X, \mathbf{R} \thickspace\sim  \thickspace \mathcal{N} (\overline{\mathbf{R}}_{*},\thickspace cov(\mathbf{R}_{*},\mathbf{R}_{*}) ) \quad , 
\]
where
	\begin{align}
	\overline{\mathbf{R}}_{*} := & \, E[\mathbf{R}_{*} | X_{*}, X, \mathbf{R}] = K(X_{*}, X) \; [K(X, X) + {\sigma}^2 I]^{-1} \; \mathbf{R} \quad , \label{eq:GPexpected}\\
	cov(\mathbf{R}_{*}, \mathbf{R}_{*}) =&\, K(X_{*}, X_{*}) -  K(X_{*}, X) \, [K(X, X) + {\sigma}^2 I]^{-1} \; K(X, X_{*}) \quad .\label{eq:GPvariance}
	\end{align}
Plugging the OS residuals into the hybrid model (\ref{eq:model}), we are able to obtain an 
{\it ex-post} probabilistic forecasting of power consumption.

The next subsection describes in detail the {\it ex-post} density forecasting technique and the last subsection the evaluation methods we consider.

\subsection{The ex-post forecasting technique and the flow diagram}

Focusing on the data structure described in the previous section, the forecast  of  middle-term daily density power consumption 
is obtained via an {\it ex-post} forecasting, a technique, introduced by \cite{HyndmanFan2010} in the power consumption sector, 
now commonly used in middle to long term power consumption forecasts \citep[see, e.g.][]{GNK2014}.

\begin{figure}[h!]
		\centering
		\includegraphics[width=.8\linewidth]{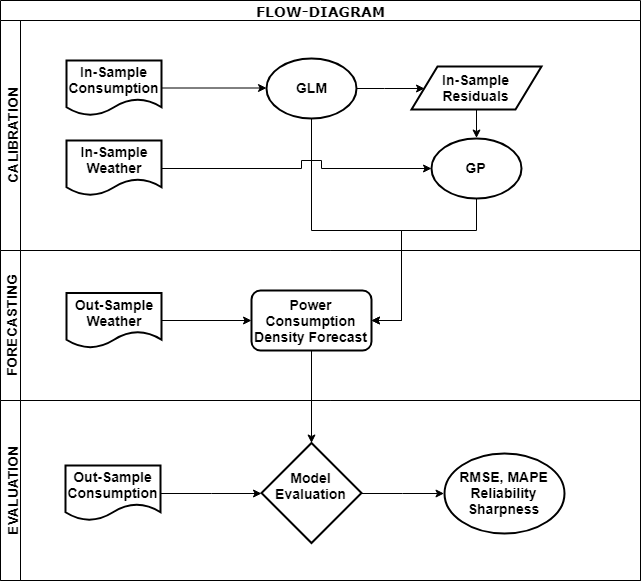}
		\caption[Flow-diagram]{Flow-diagram of the three stages of the method: calibration, forecasting and evaluation of the proposed model.}
\label{fig: flowchart}
\end{figure}

As shown in the flow diagram of Figure \ref{fig: flowchart}, the method is divided into three stages \citep[see, e.g.,][p.1144]{HyndmanFan2010}: calibration, forecasting and evaluation.

In the first stage, the GPX model (in its two components GLM and GP) is calibrated with the IS
training set, with both power and meteorological data.
GLM is calibrated through Ordinary Least Squares, while 
the calibration of Gaussian process parameters $\sigma_f$, $\sigma_l$ and ${\sigma}$  is obtained maximising the log-likelihood
\be
	\max_{\sigma_f, \sigma_l , {\sigma}} \left[ -\frac{1}{2} \mathbf{R}^\top H(X, X)^{-1} \mathbf{R} - \frac{1}{2}\ln\, \det \, H(X, X) - \frac{n}{2}\log\,2\pi\right] \, , 
\label{eq:loglik}
\en
where $H(X,X)$ is defined in (\ref{eq:covariance}).
The log-likelihood is maximised
through a Gradient descent iterative procedure with an adaptive step length \citep[see, e.g.,][and references therein]{Rasmussen06}. 

In the second stage, the density forecasting is obtained via an {\it ex-post}  forecast.
This forecast uses the weather conditions
in the OS set in order to forecast the power consumption; as well explained  by
\citet[][p.443]{GNK2014}
``assuming 
that the realisation of the meteorological covariates is known in advance (...) allows
us to quantify the performances of our model without embedding the meteorological
forecasting errors''.
The idea is that,
in order to focus on the ability of the model to describe a strong and reliable relation between daily weather conditions and power consumption,
one supposes to know perfectly the weather conditions in the OS period.

With GPX an {\it ex-post} point forecasting is straightforward. The expected value of consumption is obtained via (\ref{eq:model})
where residuals' expected value is given by (\ref{eq:GPexpected}).
The main strength of GPX is that the whole OS (conditional) distribution can be easily obtained.
OS log-consumption at day $t=n+i$, given IS consumption and weather conditions up to day $t$, is a Gaussian r.v. with
conditional mean equal to
\be
\left\{
\begin{array}{lcll}
\overline{Y}_{n+1} & = & T_{n+1} + S_{n+1} + \gamma Y_{n} + \overline{R}_{n+1} &  1=1 \; ,\\
\overline{Y}_{n+i} & = & T_{n+i} + S_{n+i} + \gamma \overline{Y}_{n+i-1} + \overline{R}_{n+i} &  2\le i\le n_{*}
\end{array}
\right.
\label{eq:cond_mean}
\en
where $\overline{R}_{n+1}$ is obtained in (\ref{eq:GPexpected}) and conditional variance
\be
\left\{
\begin{array}{lcll}
	var(Y_{n+1}) &=& var(R_{n+1}) & i =1 \; ,\\
	var(Y_{n+i}) &=& var(R_{n+i})  + \gamma^2 \; var(Y_{n+i-1}) + 2 \, \sum^{i-1}_{j=1} \gamma^j cov(R_{n+i}, R_{n+i-j})  & 2\le i\le n_{*} \, ,
\end{array}
\right.
\label{eq:var}
\en
where $cov(R_{n+i}, R_{n+i-j})$ is obtained in (\ref{eq:GPvariance}).

Equation (\ref{eq:var}) is the most relevant modeling result in this paper: 
it allows to obtain the {\it ex-post} density forecast for the proposed consumption model (\ref{eq:model}) and (\ref{eq:GP_def}). 
It extends the known formula for  autoregressive processes in presence of i.i.d. residuals 
\citep[see, e.g.,][Ch.3.2.3 p.58]{box15} to the case of interest where residuals are modeled via a GP. It can be proved via an induction method.
	

Finally, in the third stage, 
the quality of model forecast is evaluated comparing it
with the last year of realised OS consumption data. Model evaluation is realised both in terms of point consumption forecast and of reliability and sharpness of predicted densities.
These evaluation methods are described in the next subsection.

\subsection{Evaluation methods}


Besides the standard measures of point consumption forecasts as Root Mean Squared Error (RMSE) and Mean Absolute Percentage Error (MAPE), 
we provide some evaluation methods of density forecasting
that  are becoming standard in the power sector density forecasts.
Compared to point forecast that provides only  the expected value, when dealing with density forecast 
it is more difficult to value the quality of the forecast, because
we are not able to observe the realised distribution of the underlying process. 
Therefore, we cannot compare the predicted distribution to the true one, as we only have one realisation for each distribution.
The evaluation is based on two main measures: 
the \textit{sharpness} 
that verifies that 
the forecast is as tight as possible around the expected value, 
and the \textit{reliability}
that attests distribution's statistical significance.
For their detailed description we refer to \citet{NOWOTARSKI20181548}; in this subsection, we briefly summarise their main characteristics.

\bigskip

\textit{Sharpness} is measured via the Winkler score and the  pinball loss function. 
The score function proposed by \cite{winkler1972decision}, now known as the Winkler score,
 is one of the main measures for Confidence Intervals (CI) sharpness. 
Let $\hat{L}_t$ and $\hat{U}_t$ be, respectively, 
the lower and upper bounds 
for a given (central) $q$ CI, where $q$ is the CI level,
 and $y_t$ the actual consumption at time $t$,
 the Winkler score is defined as:
\[
Winkler\left(q\right):= \frac{1}{n_*} \sum^{n+ n_*}_{t=n+1}  W\left(\hat{L}_{t},\hat{U}_{t}, q; y_t\right)
\quad ,
\]
where
\[
W\left(\hat{L}_{t},\hat{U}_{t},q; y_t \right) := \begin{cases} \hat{U}_t -  \hat{L}_t, &  \quad \mbox{if }  \hat{L}_t \le y_t \le  \hat{U}_t \\ 
\hat{U}_t -  \hat{L}_t + 2(\hat{L}_t - y_t)/(1-q) , & \quad \mbox{if } y_t <  \hat{L}_t \\ 
\hat{U}_t -  \hat{L}_t + 2(y_t - \hat{U}_t)/(1-q), & \quad \mbox{if } y_t >  \hat{U}_t  \end{cases}  
 \quad .  
\]
The function $W(\cdot)$ is equal to the CI if an observation (the actual power consumption) lies inside the forecasted CI and it adds 
a penalty if the observation lies outside the CI: in this way it rewards a forecaster for a sharp (narrow) and accurate CI. 
A lower score indicates a better probabilistic forecast.

The pinball loss function is an error measure for quantile forecasts; it is the function to be minimised in quantile regression. 
Let ${\hat{Y}}_{t,q}$ be the consumption forecast at the $q^{th}$ quantile,
then the pinball loss function can be written as:
\[
 Pinball\left(q\right):= \frac{1}{n_*} \sum^{n+ n_*}_{t=n+1}  P\left(\hat{Y}_{t,q},q ; y_t\right)
 \quad ,
\]
where
\[
 P\left(\hat{Y}_{t,q},q; y_t\right):=
\begin{cases} (1-q) \, (\hat{Y}_{t,q}-y_t), & \quad \mbox{if } y_t<\hat{Y}_{t,q} \\ 
q \, (y_t-\hat{Y}_{t,q}), & \quad \mbox{if } y_t\ge\hat{Y}_{t,q} \end{cases}
 \quad . 
\]
Let us notice that 
not only the value of the pinball loss provides us a useful information but also its shape as a function of the quantile $q$.
An asymmetric pinball loss indicates that the density forecast does not reproduce with the same accuracy right and left tails of the true consumption density, 
while a symmetric pinball loss suggests that the shape of the actual distribution is forecasted adequately. 
We remind that in power consumption not only higher consumptions matter (right tail) but also lower ones that can lead to negative electricity prices. 


Both  Winkler and pinball losses are proper scoring rules \citep[see e.g.][]{Grushka2017}, 
i.e. they are minimised by the true distribution  \citep[see e.g.][]{winkler1968, gneiting2007strictly}:
this fact makes them appealing for sharpness density evaluation.

\bigskip

\textit{Reliability} refers to the statistical consistency between the density forecasts and the realised observations OS in the test set; e.g., 
if 90\% of the realised daily power consumptions fall within the 90\% CI, then this CI is said to be reliable. 

A simple and intuitive way to check model reliability from a qualitative point of view  has been shown in \cite{mori2005probabilistic}. 
For a given  nominal level $q$, one considers the (central) $q$ CI and the indicator $I_t$ that takes two values, 
$1$ if the actual consumption falls within the forecasted CI and zero otherwise, i.e.
\[
I_t = \left\{ 
\begin{array}{lll}
1 & {\rm if } \; y_t \in [\hat{L}_t, \hat{U}_t] & {\rm hit}\\
0 & {\rm if } \; y_t \notin [\hat{L}_t, \hat{U}_t] & {\rm violation}
\end{array}
\right.
\]
The empirical coverage is the OS mean of the indicator. The closer is the empirical coverage to the nominal level, the better it is.
In Section \ref{sec:results}, we show both the nominal level and the empirical coverage for several values of $q$;
in particular, we show that for GPX these two values 
result to be very close.

In order to verify that the two sets are close enough even from a quantitative point of view, it has become standard in the banking industry to run two statistical tests;
if $I_t$ is considered related to quantiles instead of CI, 
these tests correspond to the two most common risk management tests.
The first one is named 
Unconditional Coverage and tests the zero hypothesis that the empirical coverage equals the nominal level, 
i.e. $ \mathbb{P} (y_t \in [\hat{L}_t, \hat{U}_t]) = q$  \citep{Kupiec1995}.
Let $n_0$ and $n_1$ be respectively the number of zeros and ones of the indicator $I_t$,
the test is carried out in the likelihood ratio (LR) framework
\[
LR_{UC} := -2 \ln \frac{(1-q)^{n_0}\; q^{n_1} }{(1-\pi)^{n_0}\; \pi^{n_1}}
\]
where $\pi = n_1/(n_0 + n_1)$ is the empirical coverage. $LR_{UC}$ is distributed asymptotically for large $n_*$ as a $\chi^2(1)$ \citep{Kupiec1995}.

The second one is named 
Conditional Coverage and tests the alternative 
hypothesis that the ones and the zeros are clustered together in the indicator $I_t$ timeseries.
In the alternative model the time-series is modeled as a first-order Markov chain.
Let $n_{ij}$ be the number of observations with the value $i$ for the indicator $I_t$ followed by $j$ for $I_{t+1}$ and $\pi_{ij} := n_{ij}/(n_{i0} + n_{i1})$, 
the LR statistics is 
\[
LR_{CC} := -2 \ln \frac{(1-q)^{n_{00} + n_{10}}\; q^{n_{01}  + n_{11}} }{(1-\pi_{01})^{n_{00}} \;  \pi_{01}^{n_{01}} \; (1-\pi_{11})^{n_{10}} \;  \pi_{11}^{n_{11}}} \quad .
\]
This LR statistics  is distributed asymptotically as a $\chi^2(2)$ \citep{Christoffersen1998}.

\bigskip

In the following section we summarise the main results in the three steps described in \Cref{fig: flowchart}: calibration, forecasting and evaluation.

\section{Results}
\label{sec:results}	

We compare the GPX model with simple benchmarks in the power industry: the GLM (with i.i.d. residuals) and the ARX model,
 where weather conditions are introduced as linear regressors \citep[see, e.g.,][p.534]{box15}.
The three models are calibrated In-Sample (2y data from the $1^{st}$ of April $2014$ to the $31^{st}$ of March $2016$) and 
the quality of the forecast is tested in the Out-of-Sample set (1y data from the $1^{st}$ of April $2016$ to the $31^{st}$ of March $2017$).

The great majority of the studies in power is focused on price forecast \citep[see, e.g.,][for a review]{NOWOTARSKI20181548}; 
in this paper we focus on a middle-term density forecast for power consumption, where the number of studies in the literature is rather limited \citep[see, e.g.,][]{HongFan2016}.
As discussed in previous section (see Figure \ref{fig: flowchart}), 
the analysis is divided in three steps: calibration, forecasting and evaluation.

\bigskip

After a data pre-processing, the GPX model (\ref{eq:model}) and (\ref{eq:GP_def})
is calibrated In-Sample considering first the GLM and then the GP.

The data pre-processing  consists in the treatment of leap years and outliers.
We remove from the dataset February the $29^{th}$ in leap years.
Outliers may influence seasonality analysis. For this reason, they have
been removed following the same technique described in \citet{Benth2008}. 
Following that technique only one outlier has been detected corresponding to the $25^{th}$ of July 2015. It has been removed and 
the GLM is calibrated though Ordinary Least Squares. 
After estimating the GLM, the outlier has been inserted back into the time series.

\begin{figure}[h!]
	\hspace{-1.2cm}
	\includegraphics[width=1.2\linewidth]{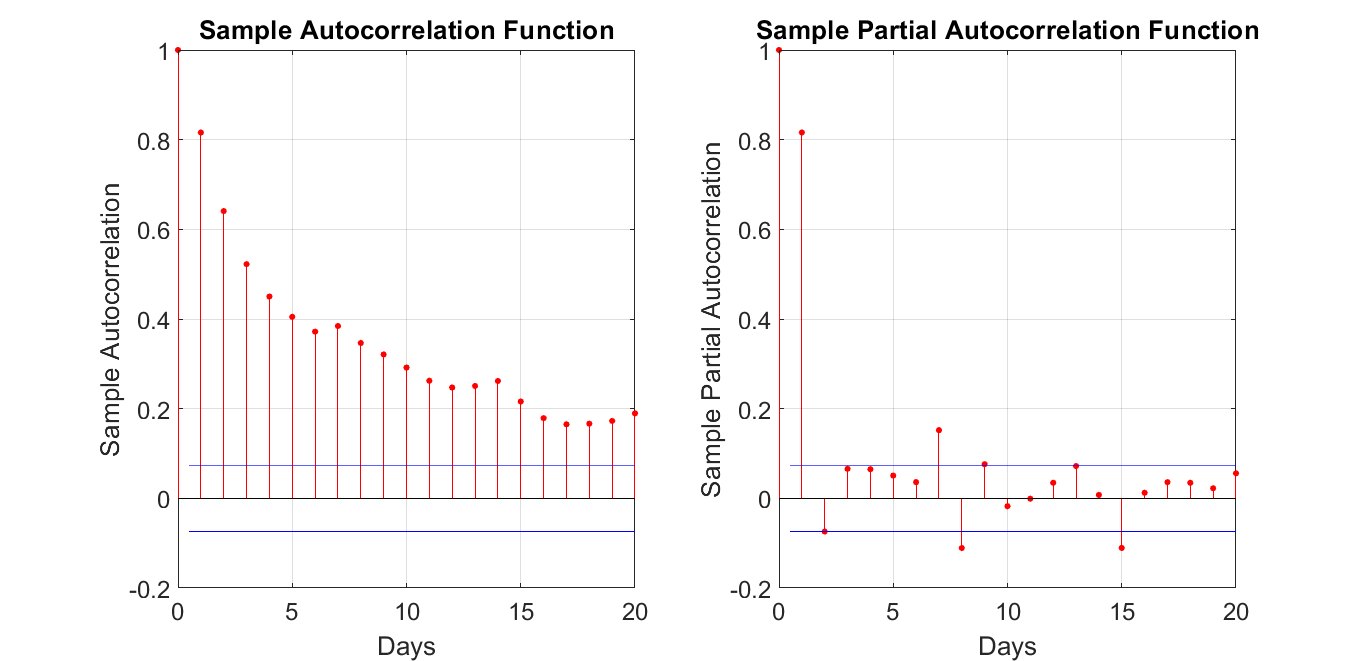}
	\caption[Autocorrelation function and partial autocorrelation function 
of log-additive deseasonalized consumption]{Autocorrelation function and partial autocorrelation function  
of seasonally adjusted consumption timeseries. We observe that an AR(1) 
	well explains the observed autocorrelation in the time-series. The horizontal lines indicate the 95\% CI.
 }
\label{fig:acf_pacf}
\end{figure}

The GLM considers both time effects and an autoregressive component, that are calibrated IS in three steps.
First, only the parameters related to the time effects are calibrated through an Ordinary Least Squares;
they include a long term trend $T_t$ and a (yearly and weekly) seasonality $S_t$. 
Second, we measure the autocorrelation and the partial autocorrelation of the residuals of this regression;
\Cref{fig:acf_pacf} highlights the necessity of a one day auto-regressive component.
Furthermore, the augmented Dickey-Fuller test refuses the null hypothesis of a unit root in favour of the alternative for the de-seasonalized process, 
ensuring stability of autoregression\footnote{An AR(1) model maximises also the BIC and AIC criteria. 
Moreover, 
the model is robust to the inclusion of local and national UK holidays.}.
Finally, we perform an IS calibration of all GLM parameters. Parameters are reported in {\Cref{tab:GLMcoeff}. 
\begin{table}[h!]             
	\centering                
	{\footnotesize \begin{tabular}{|c|c|cc|}    
\hline 
& & Estimate & (SE) \quad \\
\hline 
Intercept & $\beta_0$ & 1.17*** & (0.14) \\ 
\hline 
Trend & $\beta_1$ & -4.36e-05*** & (1.1e-05) \\
\hline 
Saturdays & $\beta_2$ & 0.022*** & (0.005) \\ 
\hline 
Sundays & $\beta_3$ & 0.047*** & (0.005) \\ 
\hline
Cos & $\beta_4$ &  0.035*** & (0.006) \\  
\hline 
Sin & $\beta_5$ & -0.103*** & (0.012) \\ 
\hline 
$AR$ & $\gamma$ & 0.819*** & (0.021) \\ 
\hline
	\end{tabular}}             
	\caption{GLM parameters calibrated IS with their standard deviation (SE). With *** we indicate
statistical significance of the parameters at 1\% significance level.}  
	\label{tab:GLMcoeff}
\end{table}        
After the calibration of the linear part of the hybrid model, GLM residuals are then calibrated via the GP described in Section \ref{sec:methodology},
maximising of the log-likelihood (\ref{eq:loglik}). 
Calibrated parameters are reported in \Cref{tab:GPXcoeff}.
\begin{table}[h!]           
	\centering                
	{\footnotesize \begin{tabular}{|c|cc|}    
			\hline 
			& Estimate & (SE) \qquad\\
			\hline 
			 $\sigma$ & 3.025e-02*** & (6.5e-05) \quad \\ 
			\hline 
			 $\sigma_f$ &  0.240*** & (3.8e-03)\\ 
			\hline
			 $\sigma_l$ &  87.8*** & (3.5) \\  
			\hline 
	\end{tabular}}             
	\caption{GP parameters calibrated IS with their standard deviation (SE). With *** we indicate
		statistical significance of the parameters at 1\% significance level. Standard errors are obtained by means of parametric bootstrapping technique \citep[see, e.g.,][]{Efron86} with 1000 samples. }  
	\label{tab:GPXcoeff}
\end{table}

\bigskip


Ex-post forecasting is straightforward with GPX: 
each density forecast at time $t=n+1, \ldots, t= n + n_{*}$ is Gaussian with conditional mean (\ref{eq:cond_mean}) and conditional variance (\ref{eq:var}).
In \Cref{fig:powerpred}, we show the OS power consumption forecasting of GPX: the continuous pink line indicates the point forecast 
while the transparent bright red indicates the 95\% confidence interval; we also show with a dot dashed green line the realised OS power consumption. 
Results look impressive: 
not only the point forecast tracks closely the realised consumption (even the spikes in winter time are tracked very closely),
but also the realised consumption falls within the $95\%$ CI in all but 15 days (95.89 \%), 
and the densities reproduce the observed behaviour of 
periods of low volatility in summer time followed by periods of high volatility in winter time.
\begin{figure*}[h!]
		\hspace{-1.2cm}
		\includegraphics[width=1.2\textwidth]{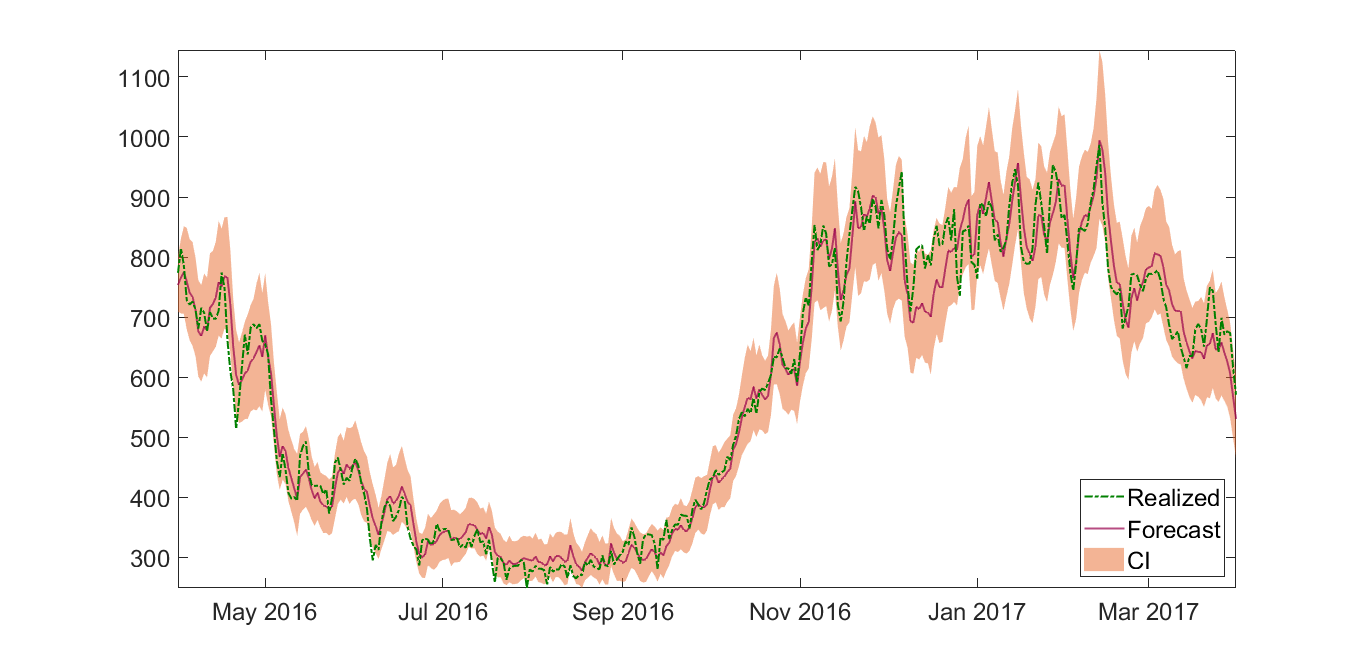}
		\caption[Power consumption density prediction.]{Realized (dot dashed green line) and expected (continuous pink line) power consumption in MWh between April 2016 to March 2017 with predicted Confidence Intervals at $95\%$ (transparent bright red).}
\label{fig:powerpred}
\end{figure*}
Let us underline that the last power consumption considered in model calibration is the $31^{st}$ of March 2016, while the forecast goes up to one later.

The quality of this forecast is the main result of the paper. 
In the remaining part of this section we provide some quantitative criteria that show the goodness
of both the point and the density forecasting.


\bigskip

For the evaluation on the model, GPX model is compared to two simple benchmark models in this field: 
the GLM and the ARX. 
The former is equivalent to a Gaussian process regression with a diagonal covariance matrix ($\sigma_f = 0$). 
The latter is an extension of GLM, in which it is introduced a linear dependency with respect to exogenous variables. 
We first consider accuracy measures for the point forecasting and then we show the results for the sharpness and reliability of the density forecasting:
these evaluation techniques have been described in Section \ref{sec:methodology}.
	

First, we value  Root Mean Squared Error (RMSE) and Mean Absolute Percentage Error (MAPE)  	
as point forecasting 
of future power consumption plays a relevant role in forecasting.  
\begin{table}[h!]
		\centering
		{\footnotesize \begin{tabular}{|c|c|c|>{\columncolor[gray]{0.9}}c|}
			\hline                                  
			& GLM & ARX & \textbf{GPX} \\                   
			\hline                                  
			RMSE & 79.24 & 46.92 & \textbf{33.84} \\        
			\hline                                  
			MAPE (\%) & 9.94 & 5.90 & \textbf{4.59} \\          
			\hline     
		\end{tabular}}
\caption{RMSE and MAPE for the three models considered. 
We observe that GPX not only presents a MAPE lower than $5\%$, 
i.e. it is considered a good forecast by practitioners, 
but also the lower absolute error indicates a more precise point forecasting in winter times, i.e. when forecasting is more relevant.}
\label{tab:sharp_errors}
\end{table}
In \Cref{tab:sharp_errors}, we compare RMSE and MAPE of the three models in the Out-of-Sample set. 
One can notice that GPX and ARX are better than GLM in terms of RMSE and MAPE, being GPX the best one. 
In particular, 
the MAPE for GPX is lower than $5\%$, 
the threshold that limits - for practitioners - a good forecast in power consumption. 
Moreover, 
a lower RMSE (almost one third lower than ARX) indicates that the GPX reduces significantly the error also in winter times, 
when the forecast is more relevant due to the higher consumption in absolute terms and the higher volatility: a behaviour we have observed in \Cref{fig:powerpred}.

\bigskip
		
Second, we consider  the analysis of sharpness.
The bottom panel in \Cref{fig:pinball_winkler} represents the Winkler score for the three models for confidence levels ranging 
from the $1^{st}$ to the $99^{th}$ percentiles. 
We observe that GPX for Winkler score is significantly lower 
than the other two models for all percentiles.
\begin{figure}[h!]
		\hspace{-1.2cm}
		\includegraphics[width=1.2\textwidth]{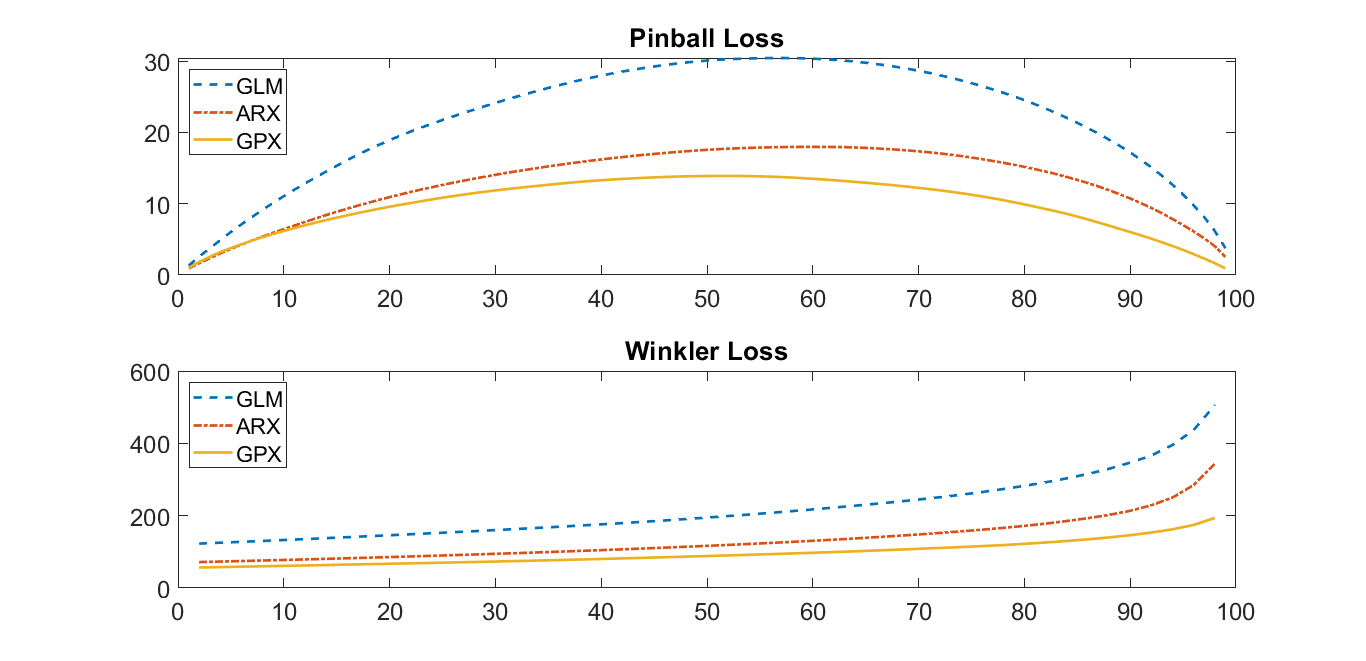}
		\caption{Pinball \& Winkler loss functions for the three models for the $1^{st}$ to the $99^{th}$ percentiles. 
We observe that not only GPX 
presents the lowest score for all percentiles (i.e. it is sharper and more accurate), 
but also that the pinball loss is more symmetric for GPX than for the other two models.}    
\label{fig:pinball_winkler}
\end{figure}

On the other hand, interesting results arise also from the analysis of the pinball loss. 
Top panel in \Cref{fig:pinball_winkler} shows pinball loss for the $1^{st}$ to the $99^{th}$ percentiles of GLM, GPX and ARX predictions.
We observe that the plot of the pinball loss provides useful information 	
not only in terms of sharpness and accuracy (the pinball loss is lower for all percentiles than the other two models and then GPX is sharper)
but also related to its symmetric shape. 
The proposed density forecasting of power consumption  is reproducing with the same accuracy both right and left tails of the actual consumption density.
	

\bigskip
		
Finally, also the analysis of reliability is presented.
 \Cref{tab:backtest} provides the backtested Confidence Intervals.
The qualitative results of this evaluation method in terms of reliability of  the proposed density forecasting look very good:
it is possible to notice that GPX backtested CI are close to the actual one, with a maximum absolute error of 2.1\% at the 90\% level. One can also notice the strong reliability of GPX at 95\% and 99\%, which are the most used nominal levels. 
Moreover, 
one can see in \Cref{fig:backtest} that the GPX Backtested CI are much closer to nominal ones for any choice of nominal level.

\begin{table}[h!]
		\centering
		{\footnotesize 	\begin{tabular}{|c|c|c|>{\columncolor[gray]{0.9}}c|}
			\hline
			Nominal level & GLM  	 & ARX 	  & \textbf{GPX}    \\ \hline
			 90\% & 76.4\% & 78.1\% & \textbf{85.5\%} \\ \hline
			 95\% & 86.0\% & 84.9\% & \textbf{91.0\%} \\ \hline
			 99\% & 94.8\% & 94.3\% & \textbf{98.4\%} \\ \hline
		\end{tabular}}
		\caption[BACKTEST MATRIX]{Backtested Confidence Intervals. We observe that the empirical coverage for GPX is very close to nominal levels. }
		\label{tab:backtest}
\end{table}
\begin{figure*}[h!]
	\centering
	\includegraphics[width=.7\textwidth]{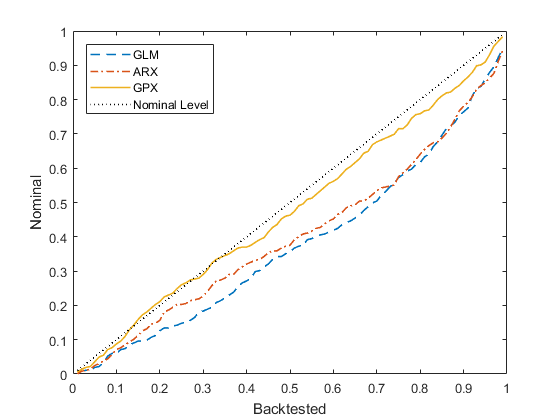}
	\caption{Backtested Confidence Intervals. We observe that the empirical coverage for GPX is very close to nominal levels.}
	\label{fig:backtest}
\end{figure*}
	
The solidity of our model can be also tested from a quantitative point of view through likelihood ratio (LR) tests. 
In particular, 
\Cref{tab:likelihoodratios} resumes 
the results of the reliability tests, standard in the backtesting of VaR in the banking industry after the introduction of Basel II: the Unconditional Coverage and the Conditional Coverage.
It is possible to notice that only GPX obtains LRs small enough to pass the tests, while they are rejected for both GLM and ARX.
\begin{table}[h!]
		\centering
		{\footnotesize 	\begin{tabular}{|c|c|c|>{\columncolor[gray]{0.9}}c|c|}
			\hline
			& GLM      & ARX     & \textbf{GPX}  & $\chi^2$ Statistic  \\ \hline
			Unconditional Coverage & 32.651 & 39.638 & \textbf{1.280} & 2.706 \\ \hline                                                     
			Conditional Coverage & 88.478 & 114.738 & \textbf{1.494} & 4.605 \\  \hline                                                     
		\end{tabular}}
		\caption[Likelihood Ratios]{Likelihood Ratios tests at 90\% level of 99\% CI. The $\chi^2$ test statistic represents the threshold over which one should reject the null hypothesis. }
		\label{tab:likelihoodratios}
\end{table}

\Cref{tab:backtest} (together with \Cref{fig:backtest}) and \Cref{tab:likelihoodratios} are the strongest results of our evaluation analysis: 
the GPX model is able to provide reliable confidence intervals over one-year time horizon for daily power consumption. 
In fact, not only we have a MAPE lower than 5\% over the whole time-horizon, 
but the nominal level and the empirical coverage appear very close for all values of $q$ considered in \Cref{tab:backtest} and shown in \Cref{fig:backtest}.
The reliability is confirmed by the LR tests;
the Conditional Coverage has shown also that the violations are not clustered in a particular period of the year, 
as it is revealed also by a direct inspection in \Cref{fig:powerpred}.
The results imply that GPX model is able to catch a very accurate relation between weather conditions and power consumption distribution.

\section{Concluding remarks}
\label{sec:conclusions}

In this paper we have introduced the GPX (cf. equations (\ref{eq:model}) and (\ref{eq:GP_def})), 
a new hybrid model for power consumption where weather conditions are included
via a simple ML technique: a Gaussian Process.

This technique allows to provide in an elementary way a density forecast
over a middle-term horizon, a forecast that is very important 
both for network reliability of power systems and for investment strategies in new plants and transmission facilities.

In terms of point forecasting we are able to predict daily power consumption with a MAPE lower than 5\% over one year (cf. \Cref{tab:sharp_errors}); 
while for what concerns density forecasting, 
we have shown in \Cref{tab:backtest}, \Cref{fig:backtest} and \Cref{tab:likelihoodratios} that is reliable, accurate and sharpe - even with a small dataset - in a detailed evaluation analysis of the results.
	
\section*{Acknowledgements}
We thank all participants to the $5^{th}$ EFI workshop in Rome (2020).
We are grateful in particular to 
F. Cordoni, P. Falbo, C. Lucheroni and P. Sabino
 for useful comments.
We thank M.Azzone for nice discussions on this topic. 


\bigskip

\section*{Abbreviations}
\addcontentsline{toc}{chapter}{Abbreviations}
\markboth{}{Abbreviations}

{\small \begin{longtable}{|c|l|}
		\hline
		$ARX$           & Auto Regressive eXogenous  model       	\\ \hline
		$GLM$           & General Linear Model       	\\ \hline
		$GP$    	& Gaussian Process                                  		\\ \hline
		$GPX$           & Gaussian Process eXogenous  model: the proposed model       	\\ \hline
		$IS$             	& In-Sample      	\\ \hline
		$LR$ 		& Likelihood Ratio \\ \hline
		$ML$		& Machine Learning \\ \hline
		$OS$             & Out-of-Sample      	\\ \hline
		$SED$ 	&  Standardised Euclidean Distance \\ \hline
\end{longtable}}

\newpage
	\section*{Notation}
\addcontentsline{toc}{chapter}{Notation}
\markboth{}{Notation}

{\small \begin{longtable}{|c|l|}
		\hline
		$Y_t$             & log-scaled power consumption time-series        	\\ \hline
		$y_t$             & realized OS log-scaled power consumption time-series        	\\ \hline
		$T_t$             & trend                       					\\ \hline
		$S_t$             & seasonality                       				\\ \hline
		$R_t$             & residual at time $t$, $t^{th}$ component of the vector $( \mathbf{R}, \mathbf{R}_{*})$       \\ \hline
		$\mathbf{R}$      & vector of IS Residuals $\in \mathbb{R}^n$         	\\ \hline
		$\mathbf{R}_{*}$      & vector of OS Residuals  $\in \mathbb{R}^{n_{*}}$         	\\ \hline
		$\{ \beta \}_{i=0,\ldots, 5}; \gamma$           & GLM parameters						\\ \hline
		$D_{Sat}(t)$          & dummy variable for Saturdays                      	\\ \hline
		$D_{Sun}(t)$          & dummy variable for Sundays                        	\\ \hline
		$n$          & IS time-series length                        	\\ \hline
		$n_{*}$          & OS time-series length                        	\\ \hline
		$\mathcal{N}(\cdot, \cdot) $     & Gaussian multivariate distribution                	\\ \hline
		$\sigma, \sigma_f, \sigma_l$        & parameters  of GP kernel      \\ \hline
		$R(x)$            & residual of Gaussian Process                         	\\ \hline
		$k(x, x')$        & kernel function of $\mathcal{GP}$                 	\\ \hline
		$\delta(x, x')$   & Kronecker delta                                   		\\ \hline
		$H(X, X)$ & covariance matrix $\in \mathbb{R}^{n \times n}$ equal to $K(X, X) + \sigma^2 \; I$        \\ \hline
		$K(\cdot, \cdot)$ & covariance matrix $\mathbb{R}^{n \times n}$ built through $k(x, x')$        \\ \hline
		I                 & Identity matrix in $\mathbb{R}^{n \times n}$           \\ \hline
		$x$               & Generic weather conditions vector                 	\\ \hline
		$\|\cdot\|$       & Standardised Euclidean Distance                                		\\ \hline
\end{longtable}}

\bigskip

\section*{References}
\bibliography{references}
	
\end{document}